\documentclass[twocolumn,showpacs,preprintnumbers,amsmath,amssymb]{revtex4}
\makeatletter
\def\@dotsep{4.5}
\makeatother
\usepackage[dvips]{graphicx}  \usepackage{amsmath}  
\usepackage{color}
\definecolor{textcolor}{cmyk}{0,0,0,1}
\definecolor{magenta}{rgb}{1,0,1}       
\definecolor{green}{rgb}{0,1,0}
\definecolor{red}{rgb}{1,0,0}

\begin{document}
\draft
\title{Interface States in Carbon Nanotube Junctions: Rolling up graphene } 
\author{ H. Santos$^1$, A. Ayuela$^{2}$, W. Jask\'olski$^3$,  M. Pelc$^3$,
and L. Chico$^1$ }   
\affiliation{  $^1$ Instituto de Ciencia de Materiales de Madrid, CSIC,  Cantoblanco, 28049 Madrid,  Spain}
\affiliation{$^2$   Centro  de F\'\i  sica  de  Materiales CSIC-UPV/EHU, 
Departamento de F\'\i sica de Materiales (Facultad de Qu\'\i micas), 
and Donostia International Physics
Center  (DIPC), 20080 Donostia, Spain} 
 \affiliation{ $^3$  Instytut  Fizyki  UMK,  Grudziadzka  5,  87-100
Toru\'n, Poland}

\date{\today}

\begin{abstract}
We  study   the  origin  of   interface  states  in   carbon  nanotube
intramolecular  junctions  between  achiral  tubes.  By  applying  the
Born-von Karman  boundary condition to an  interface between armchair-
and zigzag-terminated  graphene layers, we  are able to  explain their
number and energies.  We show that these interface states, costumarily
attributed  to  the  presence  of topological  defects,  are  actually
related  to   zigzag  edge  states,   as  those  of   graphene  zigzag
nanoribbons.  Spatial localization of interface states is seen to vary
greatly,  and   may  extend  appreciably  into  either   side  of  the
junction. Our  results give an alternative explanation  to the unusual
decay length  measured for interface states  of semiconductor nanotube
junctions, and could be further tested by local probe spectroscopies.
\end{abstract}
\pacs{}

\maketitle

%Introduction
Carbon nanotubes are  currently regarded as one of  the most promising
 materials  to  develop  future  nanoelectronics, with  an  impressive
 combination  of  robustness  and  ideal  electronic  properties.   At
 present, it  is well established  that further progress  towards real
 applications  depends  on  the  ability  to  form  junctions  between
 different   nanotubes   \cite{review}.    Recently,  the   controlled
 synthesis  of several  carbon nanotube  intramolecular  junctions has
 been  reported,   either  by  current   injection  between  nanotubes
 \cite{jin} or  by temperature changes  during growth \cite{yao_nmat}.
 These intramolecular junctions, which often present interface states,
 are typically made of topological defects arising from the connection
 between tubes of different chirality.  In fact, the interplay between
 defects and charge  transport is a central theme  of carbon nanotubes
 (CNT)  research in  the fabrication  of electronic  devices,  such as
 diodes \cite{diodes} or transistors \cite{transistors}.

 Although  interface states  are commonly  regarded as  a  drawback in
 device performance,  they may actually  provide a means  of achieving
 diode   behavior   at   the    nanoscale,   as   proposed   in   Ref.
 \cite{nlavouris}.   In any  case, transport  spectroscopy experiments
 have  shown that  interface  states  play an  important  role in  the
 behavior of CNT  junctions \cite{ej_kim03,ej_ishi04,ej_kim05}. On the
 one  hand,  Ishigami  {\it   et  al.}  studied  interface  states  in
 metal/metal CNT junctions with scanning tunneling microscopy, showing
 that interface  states extended approximately 2 nm  from the junction
 \cite{ej_ishi04}. On  the other hand,  Kim {\it et al.}  found longer
 decay  lengths for semiconductor  nanotube junctions,  with different
 values  at each  side of  the interface  \cite{ej_kim05}.  Therefore,
 understanding the physics of  CNT intramolecular junctions, for which
 interface  states  may  dominate  transport properties,  has  been  a
 subject    of   growing    activity   in    the   last    few   years
 \cite{chicoprl,junctions,chicoprb96,qds,lcwj}.
 
\begin{figure}
\includegraphics[width=0.8\columnwidth,clip]{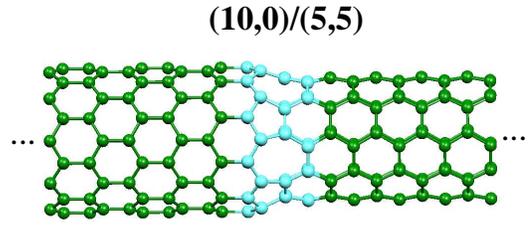}
\caption{ (Color  online) Geometry of a  $(10,0)/(5,5)$ junction, with
 the  atoms comprising  the  ring of  pentagon-heptagon  defects in  a
 different color.  }
\label{fig:fig1}
\end{figure}

The main  purpose of this letter  is to describe  interface states and
elucidate  their origin  by  studying junctions  of varying  diameter.
Specifically,  we  address  the  energy spectra  of  achiral  nanotube
intramolecular  junctions.   Zigzag/armchair  junctions  are  made  by
joining a $(2n,0)$  and an $(n,n)$ tube; this is  achieved with a ring
of  $n$ pentagon-heptagon  defect pairs.   In Fig.\  \ref{fig:fig1} we
show  a particular  example  of  this kind  of  junctions, namely  the
(10,0)/(5,5) case,  with a ring  of 5 pentagon-heptagon  (5/7) defects
forming  the union  between  the tubes.   All  calculations have  been
performed   within  the  $\pi$-electron   tight-binding  approximation
\cite{model}    and    a    Green    function    matching    technique
\cite{chicoprb96}.    We  have  recently   shown  that   for  multiple
junctions,  like  $N(12,0)/M(6,6)$  superlattices, this  approximation
yields  the electronic structure  around the  Fermi energy  ($E_F$) in
good  agreement with  the results  from  first-principles calculations
\cite{ayuela2008}.

% Model

% Results

\begin{figure}
\includegraphics[width=0.8\columnwidth,clip]{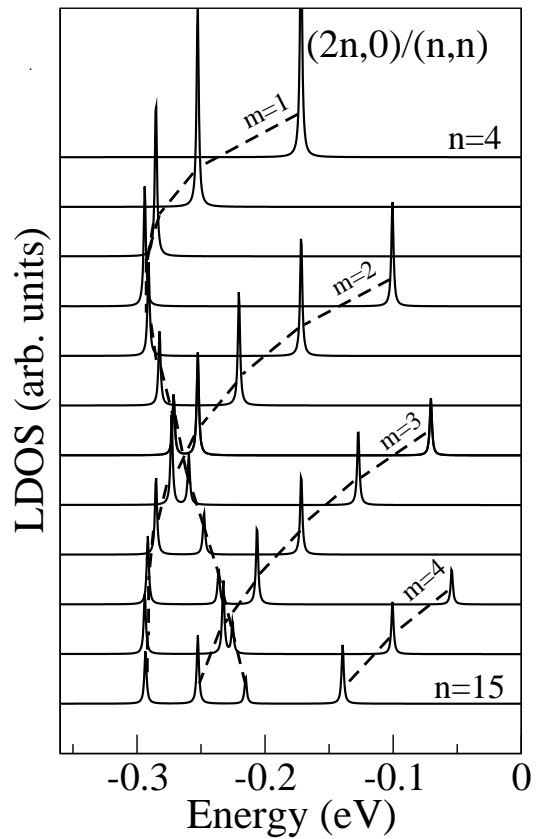}
\caption{ Local density  of states below the Fermi level  vs $n$ for a
 series  of $(2n,0)/(n,n)$  junctions. Peaks  correspond  to interface
 states.  The curves  are  arranged from  top  to bottom  in order  of
 increasing  $n$,  with  the  smallest and  largest  values  indicated
 therein.  Dashed lines are guides to the eye.  }
\label{fig:pe}
\end{figure}

In  Figure \ref{fig:pe}  we show  the local  density of  states (LDOS)
around $E_F$ evaluated at the junction, for all $(2n,0)/(n,n)$ systems
from $n=4$  to $n=15$.  The  first interface state appears  for $n=4$;
smaller junctions, such as the $(6,0)/(3,3)$ case (not shown), have no
localized states even though they  have a full ring of 5/7 topological
defects. Clearly, interface states (IS) obey a multiple-of-three rule:
when $n=3q +1$, with $q=1, 2,  ...$ a new interface state appears, $q$
being the number of such  states.  Each interface state can be labeled
by  an  integer  number   $m$,  characterizing  the  behavior  of  the
wavefunction   $\Psi_{IS}$   under  rotations   $C_n$   of  an   angle
$\phi=2\pi/n$. As  the junction is  invariant under $C_n$,  it follows
that  $C_n  \Psi_{IS}  =e^{im\phi}\Psi_{IS}$.  In  Fig.\  \ref{fig:pe}
interface states of equal $m$ are joined with a dashed line. The label
$m$ can  be viewed  as a ``discrete  angular momentum"  quantum number
\cite{chicoprb96}. The behavior of  interface states is reminiscent of
localized states  in a quantum  well system with increasing  number of
layers  \cite{qws}.   However, in  contradistinction  to quantum  well
states,   these   interface  peaks   cross   with  increasing   system
size.  Another key feature  is that  their energies  are limited  to a
narrow interval below $E_F$, specifically, between $-0.3$ and 0 eV, as
can be seen in the Figure.

\begin{figure}
\includegraphics[width=0.8\columnwidth,clip]{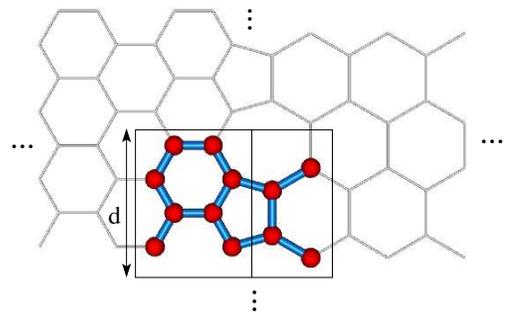}
\caption{(Color  online)  Geometry  of  the  zigzag/armchair  graphene
 junction.  The two  rectangles show  the unit  cells employed  in the
 Green function matching calculation of interface bands. }
\label{fig:geometry}
\end{figure}

Additionally,  notice that  some  interface states  at junctions  with
different $n$ have exactly the  same energies.  Such are, for example,
the  $(8,0)/(4,4)$,   the  $(16,0)/(8,8)$  and   the  $(24,0)/(12,12)$
junctions,  which  have  one  interface  state  at  $-0.172$  eV,  the
$(12,0)/(6,6)$  and the  $(24,0)/(12,12)$  junctions, with  one IS  at
$-0.285$ eV.   The coincidence in  energies for some  interface states
and  the regularity in  their appearance  point towards  their folding
origin.  Thus, we  have turned  to a  system closely  related  to this
series of nanotube junctions: a semiinfinite zigzag graphene joined to
an   armchair-terminated   one,   yielding   an   infinite   line   of
pentagon-heptagon  topological defects  as interface  between  the two
graphene edges.  The geometry  of this graphene  junction is  shown in
Fig.\ \ref{fig:geometry}. In  the same way that a  perfect nanotube is
made by rolling  up a graphene sheet, a  carbon nanotube junction like
those described above  can be obtained by rolling up  a strip of these
matched semiinfinite graphenes.
 
\begin{figure}
\includegraphics[width=0.8\columnwidth,clip]{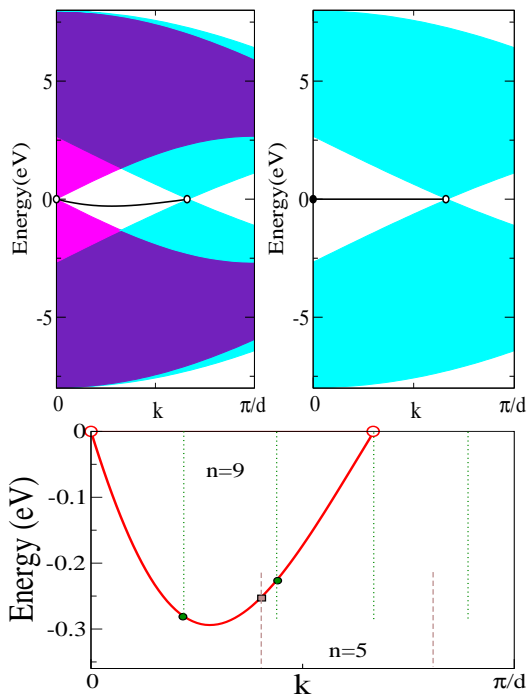}
\caption{ Left  panel: Interface band of  the zigzag/armchair graphene
junction, with the corresponding projected graphene bulk bands at this
interface.  Right  panel: Edge band of  the zigzag-terminated graphene
with  the graphene  bandstructure projected  at this  surface.  Bottom
panel: Zoom of the interface band, with quantization lines showing the
interface  energies for  the $n=5$  (dashed) and  the  $n=9$ junctions
(dotted).  Symbols mark  the corresponding  interface  state energies.
Open circles at the interface  band ends stress that these energies do
not belong to the band but, rather, to the bulk continua.  }
\label{fig:gis}
\end{figure}

The bandstructure  of the graphene armchair/zigzag  interface is shown
in the left panel of Fig. \ref{fig:gis}, along with the projected bulk
bands  at this interface;  the right  panel depicts  the edge  band of
zigzag-terminated graphene  with the corresponding  projected graphene
bulk bandstructure  at this surface.  The interface band shown  in the
left panel  spans from  $\Gamma$ to  2/3 of the  positive part  of the
Brillouin  zone.  Note that  just  at the  edge  points  there are  no
interface states, because they belong to the bulk of the armchair- and
the  zigzag-terminated graphene  respectively. The  graphene interface
band spans from $-0.3$ eV to  0 eV, comprising the energy range of all
the  nanotube  interface  states.  Rolling up  the  graphene  junction
amounts to imposing Born-von Karman boundary condition to the graphene
interface band. This determines the quantization rule
\begin{equation}
k=\frac{2\pi m}{nd}, \;\;\; m=0,...,n-1, 
\label{qrule}
\end{equation}
where $d$ is the length of the repeat unit along the interface and $n$
is the  number of repetitions  to give a $(2n,0)/(n,n)$  junction. The
index  $m$ is  the same  ``discrete angular  momentum"  label formerly
introduced.  The allowed $k$ values give the nanotube interface states
shown  in  Fig.  \ref{fig:pe},  as demonstrated  graphically  for  two
examples,   namely  $n=5$   and  $n=9$,   in  the   bottom   panel  of
Fig. \ref{fig:gis}.  The energies obtained by this  rule exactly match
those  obtained in  the nanotube  junction calculations,  collected in
Fig.   \ref{fig:pe}.  The   multiple-of-three   periodicity  is   thus
understood, due to the length of the BZ portion in which the interface
graphene   band   exists,  {\it   i.e.},   2/3   of  its   irreducible
part. Furthermore, within the model  employed, it is now clear why for
$n<4$ there  are no interface states in  the $(2n,0)/(n,n)$ junctions:
in these  cases, quantization  lines touch the  edges of  the graphene
interface  band, and these  end points  do not  actually belong  to it
because   they  are  in   the  zigzag   and  armchair   graphene  bulk
continua. Finally, the appearance of interface states with exactly the
same energies  is simply explained  by the quantization rule  given in
Eq. (\ref{qrule}).

To  understand  the  origin  of  the  interface  band  found  for  the
zigzag/armchair graphene junction,  we have analyzed the corresponding
graphene   free   ``surfaces",   the   armchair-terminated   and   the
zigzag-terminated semiinfinite  graphenes \cite{fujita96}.  No surface
bands appear  in the gap of the  armchair-terminated graphene, whereas
for  the   zigzag-terminated  one,  shown   in  the  right   panel  of
Fig. \ref{fig:gis}, a flat band at  0 eV spans from $\Gamma$ to 2/3 of
the  irreducible  part  of  the   BZ.   For  the  purposes  of  direct
comparison,  we  use   the  same  unit  cell  as   for  the  interface
calculation, which is doubled with respect to the one usually employed
for zigzag geometry.  Note that the $k=0$ state belongs to the surface
band,  given  that it  is  in  the bulk  gap;  this  explains why  all
semiinfinite $(2n,0)$ zigzag nanotubes have a ``surface" edge state at
0 eV. Joining the zigzag edge  graphene to the armchair one breaks the
electron-hole  symmetry  due  to   the  mixing  of  the  two  graphene
sublattices,  combing  the surface  band  and  moving  it to  negative
energies, as depicted  in the left panel of  Fig. \ref{fig:gis}. Thus,
the  armchair-edge graphene  acts  as an  external  potential for  the
states of  the zigzag-terminated graphene, bending  down the interface
band.   The  zigzag  edge  nature  of  the  interface  band  shown  in
Fig. \ref{fig:gis} is thus demonstrated,  as well as that of interface
states in zigzag/armchair junctions of tubes, which we have shown here
to originate from edge states, as those found in graphene nanoribbons.
This finding has  implications for the analysis of  other defects such
as vacancies, and even  substitutional atoms in nanotubes or graphene,
which  have been shown  to yield  an effective  edge in  the hexagonal
carbon lattice \cite{gsb08}.

In order to test the  robustness of our tight-binding results, we have
performed   {\it   ab   initio}   calculations   of   $4(2n,0)/4(n,n)$
superlattices  (SL)  using  the  same  method  and  parameters  as  in
Ref.  \cite{ayuela2008}.  Introducing  another  junction and  imposing
periodic  boundary  conditions  induces  significant  changes  in  the
electronic structure,  but by comparison to tight  binding results and
checking the  wavefunction symmetry and spatial  distribution, we have
successfully    identified   interface    states   \cite{ayuela2008c}.
Specifically, we  have checked that  there are no interface  states in
the $(6,0)/(3,3)$ system,  whereas one IS per junction  appears in the
$(8,0)/(4,4)$  and  the  $(12,0)/(6,6)$  cases,  and  two  states  per
junction  appear in the  $(14,0)/(7,7)$ system.   For the  time being,
further  studies for  lattices with  a  larger number  of defects  are
beyond  our  computational  capabilities.  Hitherto, {\it  ab  initio}
calculations and tight-binding results fully agree as to the number of
interface states in these achiral junctions.

\begin{figure}
\includegraphics[width=0.8\columnwidth,clip]{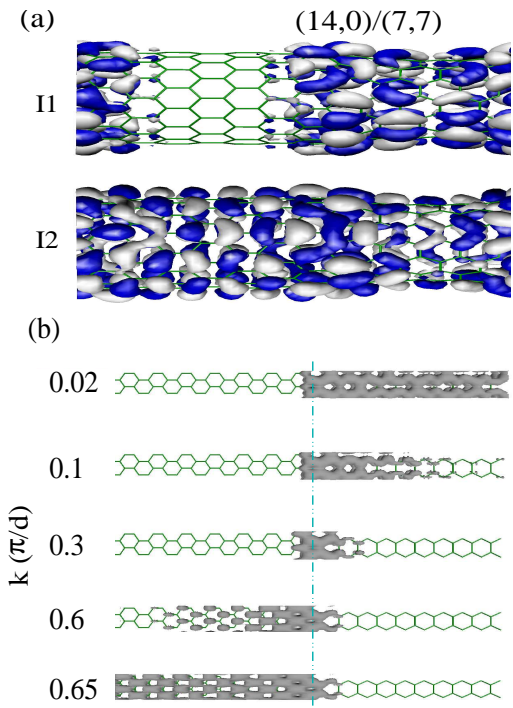}
\caption{(Color  online)   (a):  Two  examples   of  wavefunctions  of
 interface  states  belonging  to  the  $4(14,0)/4(7,7)$  superlattice
 calculated {\it ab initio}.   (b): Electron density of several states
 belonging  to  the interface  band  of  the graphene  armchair/zigzag
 junction, calculated with a $\pi$-orbital tight-binding model. }
\label{fig:gwf}
\end{figure}

 We have  chosen a pair of  interface states belonging  to the largest
system  calculated   by  {\it   ab  initio}  techniques,   namely  the
$4(14,0)/4(7,7)$  SL,  to   show  their  spatial  distribution.  Their
wavefunctions are  shown in Fig. \ref{fig:gwf}  (a).  The lowest-lying
interface  state, labeled I1,  is mainly  localized at  the interface,
spreading towards the armchair  side.  This behavior was also observed
in  the  interface states  of  $(12,0)/(6,6)$  SLs and  $(10,0)/(5,5)$
junctions \cite{ayuela2008c,nlavouris}.  But, surprisingly, the second
interface  state (labeled  I2)  spreads from  the  interface into  the
zigzag part.  To understand these disparate behaviors, we turn back to
the  graphene junction.  Fig. \ref{fig:gwf}  (b) depicts  the electron
density  of  several  graphene  interface states  with  different  $k$
values; states are labeled with the corresponding $k$ value in $\pi/d$
units. When moving  from $\Gamma$ to the interface  band edge at $2/3$
of the BZ, the wavefunction  localization changes from the armchair to
the  zigzag side;  for $k$  at the  band minimum  the  wavefunction is
mainly  localized at the  junction.  This  explains why  the interface
state of  the $(12,0)/(6,6)$ junction,  which stems from  the graphene
$k=1/3$  state,  is  rather  localized  at the  interface.  Thus,  the
junctions  with  sufficiently   large  diameter  will  have  different
interface  states spreading at  opposite sides  of the  interface, but
pinned at the carbon ring made of 5/7 topological defects.

 Our results  provide an alternative explanation to  the unequal decay
lengths found in semiconductor nanotube junctions, as well as to their
large  value compared  to metallic  systems \cite{ej_ishi04,ej_kim05}.
The  coexistence  in  the  same  nanotube  of  interface  states  with
dissimilar  spatial  localization could  be  demonstrated by  scanning
tunneling    microscopy     and    spectroscopy,    as     in    Refs.
\cite{ej_kim03,ej_ishi04,ej_kim05}.  The  fact that CNT  junctions may
have  several interface  states with  different  spatial localizations
opens  a way  for new  device design  based on  their characteristics.
Choosing a  CNT of appropriate  diameter, states with  quite different
spatial localization  can be accessed by  applying different voltages,
allowing for  switch operation.  Furthermore, due  to interface charge
localization and  redistribution CNT  junctions can act  as chemically
active sites. Actually, doping  the interfaces may induce a structural
reconstruction  and alter their  symmetry, thus  dramatically changing
the electronic properties because of  the Fermi level proximity to the
IS.

Finally, although  we have focused on junctions  between achiral tubes
and  found that  their interface  states have  zigzag edge  origin, we
would like to note that differences in chirality of joined tubes plays
a role.  For example, a  zigzag (8,0)/(7,1) junction has  no interface
states, while  the (8,0)/(5,3)  junction has two  \cite{chicoprl}.  We
have chosen  to study junctions between tubes  with maximum difference
in chiral  angles. The non-trivial role of  chirality deserves further
exploration; but in  any case, our present results  suggest that IS in
chiral systems will also have edge origin.

In summary, we have explored  the nature of interface states in carbon
nanotube junctions,  focusing on achiral  systems. We have  shown that
these states, usually  attributed to the pentagon-heptagon topological
defects,    are   actually    due   to    the   zigzag-edge-terminated
nanotube.  Topological   defects  break  the   electron  symmetry  and
consequently make these  states energy-dependent. Furthermore, we have
related these interface states  of nanotube junctions to the interface
band appearing  in a  graphene zigzag/armchair junction.   By applying
the Born-von Karman boundary condition to the graphene interface band,
we  have derived  the energies  and number  of the  nanotube interface
states,  obtaining  complete   quantitative  agreement  with  the  CNT
junction calculations.  Our results give a new vision on the nature of
CNT interface states  and have implications in other  systems, such as
graphene vacancies or substitutional impurities.

L. C. acknowledges  helpful discussions with J. I.  Cerd\'a. This work
has  been  partially  supported  by  the  Spanish  DGES  under  grants
MAT2006-06242 and  FIS2007-66711-C02-C01 and Spanish  CSIC under Grant
PI  200860I048. W.  J. and  M. P.  acknowledge financial  support from
Polish LFPPI.

\end{document}